\documentclass[pre,twocolumn,epsfig,rotate,noshowpacs]{revtex4}
\usepackage{graphicx}
\usepackage{amsmath}
\usepackage{dcolumn}
\usepackage{epsfig}
\usepackage{bm}
\begin{document}

\title{Anomalous temperature dependent heat transport in one-dimensional momentum-conserving systems with soft-type interparticle interaction}
\author{Daxing Xiong}
\email{phyxiongdx@fzu.edu.cn}
\affiliation{Department of Physics,
Fuzhou University, Fuzhou 350108, Fujian, China}

\begin{abstract}
We here numerically investigate the heat transport behavior in a one-dimensional lattice with a soft-type (ST) anharmonic interparticle interaction. It is found that with the increase of system's temperature, while the introduction of ST anharmonicity softens phonons and decreases their velocities, this type of nonlinearity like its counterpart of hard type (HT), can still not be able to fully damp the longest wave phonons. Therefore, an anomalous temperature dependent heat transport with certain scaling properties similarly to those in the Fermi-Pasta-Ulam like systems with HT interactions can be seen. Our detailed examination from simulations well verify this temperature dependent behavior.
\end{abstract}
\maketitle

\section{Introduction}
As one of the fundamental topics closely related to the concepts of nonlinearity and
irreversibility in statistical mechanics~\cite{StatiticalM}, heat
transport in one-dimensional (1D) systems has attracted considerable
interest in recent
years~\cite{Lepri_Report,Dhar_Report,Lepri_Book,Meeting_2012}. In
this context, one of the central issues is the validity or breakdown
of Fourier's law, which states, the heat flux $\textbf{\emph{J}}$ is
proportional to the temperature gradient $\nabla T$:
$\textbf{\emph{J}}=- \kappa \nabla T$, with $\kappa$ the heat
conductivity assumed to be a size-independent constant. Generally,
it is now well accepted that for 1D anharmonic systems with
conserved momentum, Fourier's law is not valid, namely $\kappa$
is not a constant but diverges with the system size $L$ in a power
law $\kappa \sim
L^{\alpha}$~\cite{Lepri_Report,Dhar_Report,Casati,PRL1997,Grassberger2002,PRL2002,Olla2006,PRE2015,PRL2007}.
The exponent $\alpha$ ($0 \leq \alpha \leq 1$) is believed to following
some universality
classes~\cite{Lepri_Report,Dhar_Report,Grassberger2002,PRL2002,Olla2006,PRL2007,Hydrodynamics-1,Hydrodynamics-2,Levywalks-Rev2015,Levywalks-PRL2011,Popkov},
which was supported by some
theories~\cite{PRL2002,Olla2006,Hydrodynamics-1,Hydrodynamics-2,Levywalks-Rev2015,Levywalks-PRL2011,Popkov}
and numerical
simulations~\cite{Grassberger2002,PRL2007,Levywalks-PRE2014a,Levywalks-PRE2014b},
and debated by some other
studies~\cite{Xiong-1,Xiong-2,Xiong-3,Daswell-1,Daswell-2,Daswell-3,Hurtado}.
While it should be noted that if some other factors, such as
the asymmetric interactions~\cite{Asy-1,Asy-2,Asy-3,Asy-4,Asy-5},
the systems close to the integrable limit~\cite{Benenti}, the
pressure~\cite{Pressure-1,Pressure-2}, and the multi-well
interparticle
potential~\cite{Multi-1,Multi-2,Multi-3,Multi-4,Multi-5,Xiong-4,Xiong-5} are included, whether the Fourier's law is still valid/invalid and what are
the underlying mechanics, the verification remains in progress.

It is thus necessary to check the heat transport behavior including more
complicated factors to seek general conclusions. In the present
work we therefore consider a momentum-conserving system with the
soft-type (ST) interparticle interaction, which is a new factor that has
not yet been fully taken into account [compared with hard-type (HT)
interactions]. Our main finding is that, similarly to the Fermi-Pasta-Ulam-$\beta$ (FPU-$\beta$)
systems with HT anharmonicity, an anomalous heat transport will be observed at all temperatures. Via detailed simulations we also explore the possible microscopic mechanism. We show that with the increase of system's temperature, the ST interaction induces a
special type of nonlinearity, which softens phonons and reduces
their velocities; while these unusual effects cannot qualitatively
change the heat transport and its scaling behavior. A careful
analysis of system's momentum spread and phonons spectra
indicates an incomplete damping process of phonons very similarly to
those exhibited in FPU-$\beta$ systems
with HT anharmonicity. This may be the mechanism for the anomalous temperature
dependent heat transport behavior observed here.

The rest of this work is organized as follows: In Sec. II we
introduce the model and compare the ST anharmonic interaction with the potentials of harmonic and
FPU-$\beta$ (with HT interaction) systems. Section III describes the simulation method. We use the equilibrium
correlation simulation method~\cite{Zhao2006,Chen2013} to get the heat spreading information with temperatures, from which our main results on heat transport and its scaling property are presented in Sec. IV.
Section V is devoted to the underlying mechanisms. For such purpose we investigate the system's momentum spread and examine the phonons spectra to explore phonons' damping information. Finally, a summary is given in Sec. VI.
\section{Model}
\begin{figure}
\vskip-.2cm \hskip-0.4cm
\includegraphics[width=8.cm]{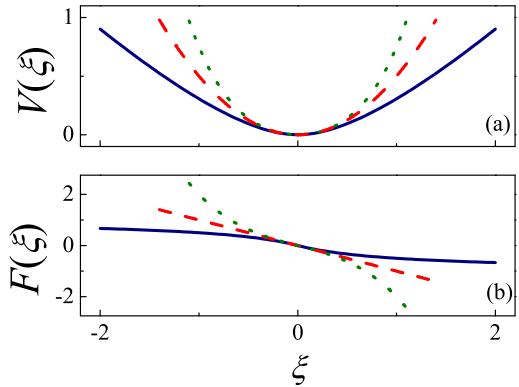}
\vskip-0.4cm \vskip-0.3cm \caption{\label{Fig1}(Color online) The ST anharmonic interparticle
potential [Eq.~\eqref{STPotential}, solid] (a) and its associated force (b). For comparison we also plot the potentials of the harmonic (dashed) and
FPU-$\beta$ (dotted) systems.}
\end{figure}

We consider a 1D many-particle ($L$ particles) momentum-conserving
lattice with Hamiltonian
\begin{equation} \label{Hamiltonian}
H= \sum_{k=1}^{L} p_{k}^2/2 + V(r_{k+1}-r_k),
\end{equation}
where $p_k$ is the $k$-th particle's momentum and $r_{k}$ its
displacement from equilibrium position. Note that here both the averaged distance between particles and the lattice constant are set unity, thus the number of particles is equal to the system size $L$. The interparticle potential
takes a type of soft anharmonicity~\cite{Soft-1,Soft-2,Soft-3}
\begin{equation} \label{STPotential}
V(\xi)= \left| \xi \right| - \ln \left(1+ \left| \xi \right|\right),
\end{equation}
which is plotted in Fig.~\ref{Fig1}(a) and also compared with the harmonic [$V(\xi)=\xi^2/2$] and
FPU-$\beta$ [$V(\xi)=\xi^2/2+\xi^4/4$] potentials.

Figure~\ref{Fig1}(b) plots the associated forces defined by $F(\xi)=-\partial V(\xi) /\partial
\xi$. As can be seen, opposite to the FPU-$\beta$ system with HT anharmonicity, the ST interaction has a restoring
force always less than the harmonic force. That is why we
call it ST anharmonicity, a special feature of the system. It induces
an unusual energy dependent frequency~\cite{Soft-1}, which has been suggested to strongly modify the distribution, intensity, and
mobility of the thermal fluctuations, resulting in a quite different
transition dynamics of the underlying activated
process~\cite{Soft-3}. In the research field of discrete breathers (DBs), such ST anharmonicity can also support a different amplitude dependent property of DBs' frequency, opposite to that induced by HT anharmoncity~\cite{DBs}. Motivated by these microscopic evidences, we here aim to explore how this ST anharmonicity would play the roles in heat transport.
\section{Method}
As mentioned, to identify the heat transport behavior and its scaling property,
we here use the equilibrium correlation method~\cite{Zhao2006,Chen2013}. For the special ST anharmonicity,
it can be expected that the time scale to ensure the system relaxed
to the nonequilibrium stationary state would be quite longer than
the usually considered FPU-$\beta$ systems with HT anharmonicity. This would be why the traditional
simulation methods, such as the direct nonequilibrium molecular
dynamics simulations~\cite{PRL1997} and the approach based on Green-Kubo formula~\cite{PRL2002} have not yet been used to study such a system, and thus the heat transport
law here is still unclear.

The equilibrium correlation method~\cite{Zhao2006,Chen2013} employs the following normalized
spatiotemporal correlation function of system's heat energy
fluctuations to explore the heat spreading information
\begin{equation}
\rho_{Q} (m,t)=\frac{\langle \Delta Q_{j}(t) \Delta Q_{i}(0)
\rangle}{\langle \Delta Q_{i}(0) \Delta Q_{i}(0) \rangle},
\end{equation}
where $m=j-i$; $\langle \cdot \rangle$ represents the spatiotemporal average; $i$ and $j$ denote the label of bins, this is because in hydrodynamics theory, the heat energy density should be defined by a function of space rather than the lattice site. Viewing this fact, we set the number of particles in the $i$-th bin to be $N_i=L/b$, where $b$ is the total number of the bins. Under this setup, in each bin one can compute the energy $E_i(t)$, particle $M_i(t)$ and pressure $F_i(t)$ densities by summing the corresponding single particle's densities $E(x,t)$, $M(x,t)$ and $F(x,t)$ at the site $x$ and time $t$ within the bin. The heat energy density in the $i$-th bin then is $Q_i(t)\equiv
E_i(t)-\frac{(\langle E \rangle +\langle F \rangle) M_i(t)}{\langle M
\rangle}$~\cite{Forster,Liquid} and its fluctuation $\Delta Q_{i}(t)\equiv Q_i(t)- \langle Q_i \rangle$ can be straightforwardly obtained. Clearly, this definition suggests that heat energy in one bin is closely related to the associated energy and particle densities under a internal averaged pressure $\langle F \rangle$ ($\equiv 0$, since the potential here is symmetric and the averaged distance between particles is set identical to the lattice constant).

In order to understand the underlying picture, we also study the momentum spread via the momentum correlation function
\begin{equation}
\rho_{p}(m,t)=\frac{\langle \Delta p_{j}(t) \Delta p_{i}(0)
\rangle}{\langle \Delta p_{i}(0) \Delta p_{i}(0) \rangle}.
\end{equation}
Similarly to the definition of $\rho_{Q} (m,t)$, here $\Delta p_{i}(t)\equiv
p_i(t)- \langle p_i \rangle$ denotes the momentum fluctuation.

To calculate the correlation functions, the system is first thermalized to the focused
temperature by using the stochastic Langevin
heat baths~\cite{Lepri_Report,Dhar_Report} for a long enough time ($>10^7$
time units of the models). This should be taken from properly assigned initial random
states. Then the system is evolved in isolation by using the Runge-Kutta algorithm of $7$th to $8$th order with a time step $h$ for deriving
the correlation information. We use the size of ensemble about $8 \times 10^9$.

We consider a wide range of temperatures from $T=0.00075$ to $T=1$. For each temperature, we set the chain's size of $L=2001$, which allows a fluctuation of heat located at the center to spread along the system for a long time at least up
to $t =600-900$. Under this setup, we apply the periodic
boundary conditions, fix the bin's number $b \equiv L/2$ (the choice of $b$ has been verified
not to affect the final results).

The difficulty of simulating the correlations for ST anharmonic systems lies in the case of high temperatures. This is 
because due to the phonons' softening (shown below), a higher $T$ may need longer time to ensure the system relaxed to the stationary state and also require higher precision of integration with a smaller time step $h$, which will cost many computing resources for the calculations. Therefore, the highest temperature considered is $T = 1$ and different time step $h$ are used for different temperatures, i.e., for low temperatures, $h=0.05$ is always adopted; while for the temperatures higher than $T=0.5$, we set $h=0.02$, 
which has been verified to be small enough for the system to evolve under satisfactory precision.
\section{Heat spread and its scaling}
\begin{figure}
\begin{center}
\vskip-.2cm \hskip-0.4cm
\includegraphics[width=8.8cm]{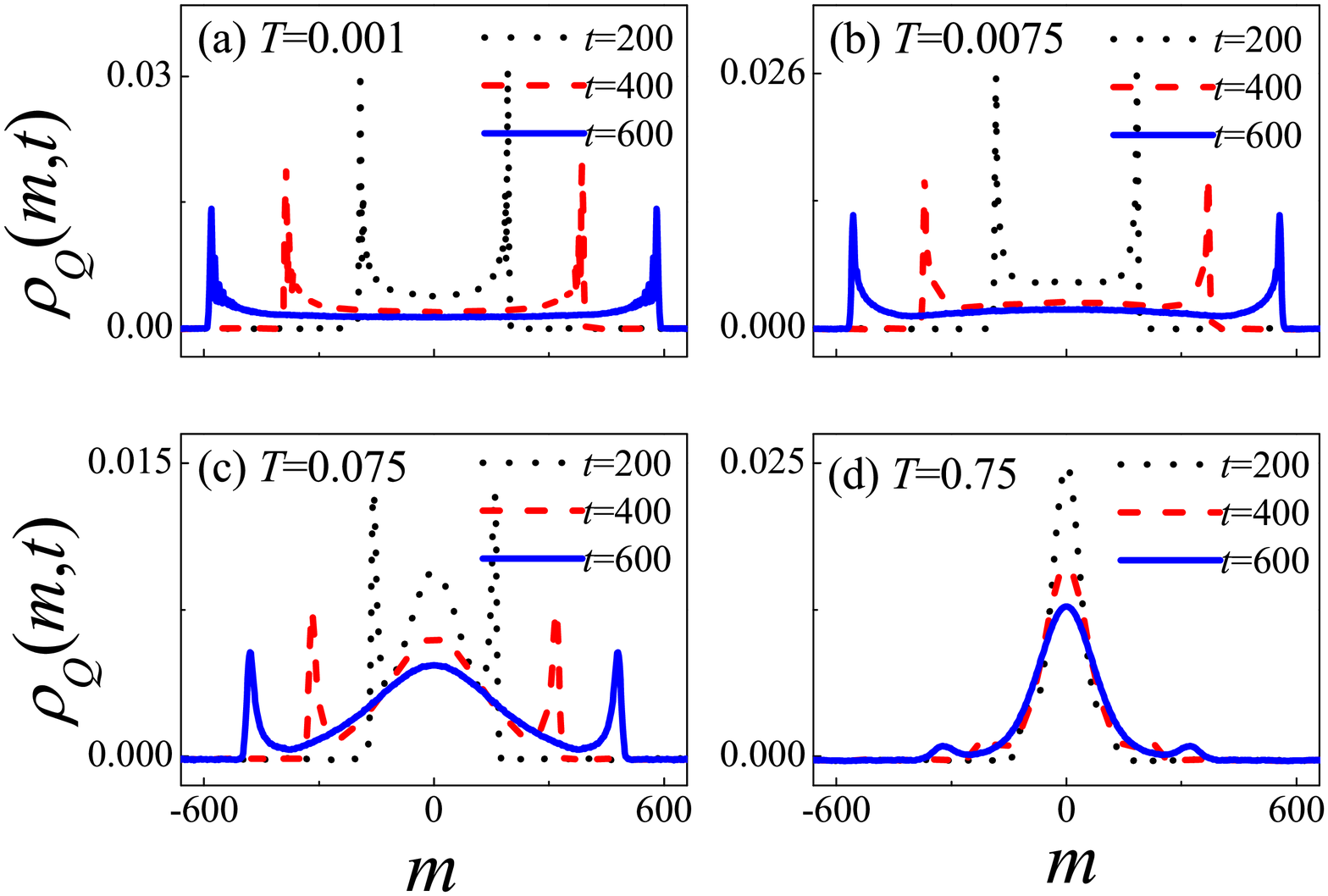}
\vskip-0.5cm \caption{\label{Fig2} (Color online) Profiles of $\rho_Q(m,t)$ for three long time $t=200$ (dot), $t=400$ (dash) and $t=600$ (solid) under temperatures $T=0.001$ (a);
$T=0.0075$ (b); $T=0.075$ (c) and $T=0.75$ (d), respectively.}
\end{center}
\end{figure}

Now let us see the results of heat spread. In Fig.~\ref{Fig2} we plot the profiles of
$\rho_Q(m,t)$ for three typical time (here the time up to $t=600$ is used for example, such choice is due to the fact that employing a longer time's result will cause the side peaks of the profiles hard to identify for high temperatures). Four temperatures $T$, from low to high, are
employed to explore the temperature dependent behavior. As can be seen, with the increase of
$T$, the profile of $\rho_Q(m,t)$ is changed from a U shaped~\cite{Ushape} to L\'{e}vy walks~\cite{Levywalks-Rev2015,Levywalks-PRL2011} density, especially that the central
parts of the profile become more and more localized. The U shape here shows slight difference with
the usual density in harmonic chain~\cite{Ushape}, i.e., the front parts exhibit some oscillations, which may be caused by the unusual nonlinearity induced by ST anharmonicity under low temperatures. The L\'{e}vy walks profiles under high temperatures can be phenomenologically understood from the single particle's L\'{e}vy walks theory considering velocity fluctuations~\cite{VelocityFluctuation}.

\begin{figure}
\begin{center}
\vskip-.2cm \hskip-0.4cm
\includegraphics[width=8.8cm]{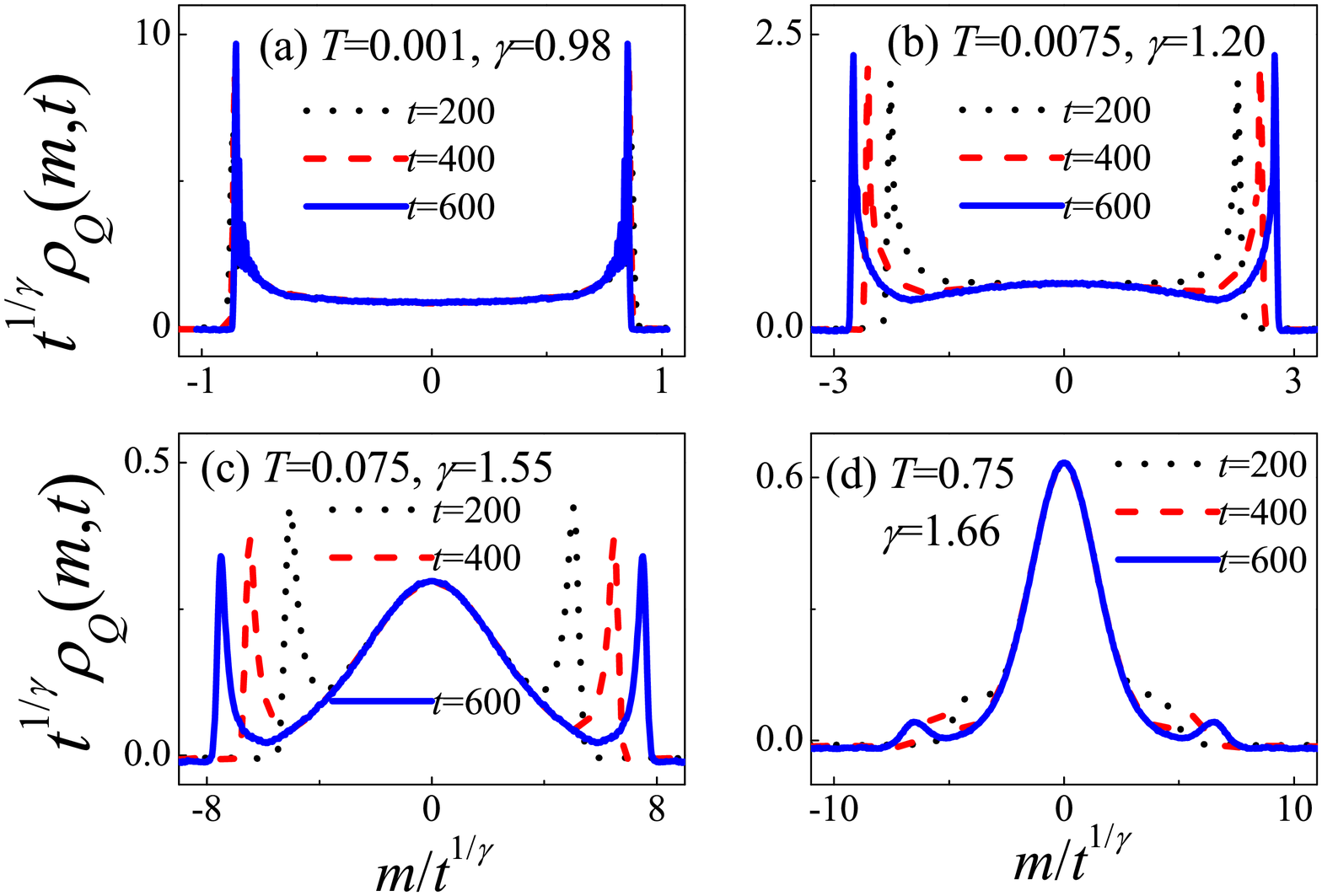}
\vskip-0.5cm \caption{\label{Fig3}(Color online) Rescaled $\rho_{Q}(m,t)$, the focused temperatures here are the same as those in Fig.~\ref{Fig2}.}
\end{center}
\end{figure}

In view of the coincidence to the L\'{e}vy walks profiles, one then can perform a scaling analysis to $\rho_Q(m,t)$ by using the following scaling formula~\cite{Levywalks-Rev2015,Levywalks-PRL2011} 
\begin{equation} \label{scaling}
t^{1/\gamma} \rho_{Q} (m,t) \simeq \rho_{Q} (\frac{m}{t^{1/\gamma}},t).
\end{equation}
Note that this scaling law only applies to the central parts if the underlying diffusion process is superdiffusive ($1<\gamma<2$), while for the ballistic ($\gamma=1$) and normal diffusive ($\gamma=2$) transport, it is valid for all of ranges~\cite{Levywalks-Rev2015,Levywalks-PRL2011}. For the formula applying to high dimensions, one can refer to a recent work on two-dimensional L\'{e}vy walks~\cite{Levy2D}. 
The rescaled profiles under formula~\eqref{scaling} are shown in Fig.~\ref{Fig3}, which then enable us to identify some space-time scaling
exponents $\gamma$ for characterizing the heat spreading behaviors. Since $\gamma=1$ and $1<\gamma<2$
correspond to the ballistic and super-diffusive heat transport,
respectively, from Fig.~\ref{Fig3} now it is clear that U shape shown at low temperatures [$\gamma=0.98$ close to $1$, see Fig.~\ref{Fig3}(a)] indicates the ballistic heat transport, while the L\'{e}vy walks density under high temperatures implies the super-diffusive behavior [Fig.~\ref{Fig3}(d), $\gamma=1.66 > 1$].
For the ballistic regime the whole density can be perfectly scaled by formula~\eqref{scaling}, both for the central and front parts, while in the super-diffusive case, only the central parts are available, which might correspond to the bi-linear scaling property of L\'{e}vy walks model~\cite{Bilinear,Bilinear-3}.
\begin{figure}
\vskip-.2cm \hskip-0.4cm
\includegraphics[width=8.cm]{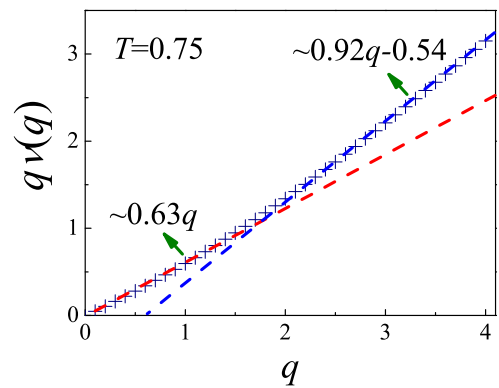}
\vskip-0.5cm \caption{\label{Fig4}(Color online) Exponents $q \nu(q)$ versus $q$ for indicating the bi-linear scaling behavior. Here we take the case of $T=0.75$ for example.}
\end{figure}

To demonstrate the bi-linear scaling behavior in the superdiffusive regime, following Refs.~\cite{Bilinear,Bilinear-3} we use $\rho_Q (m,t)$ to calculate the $q$ order momentum, i.e., $\langle |m(t)|^q \rangle = \int_{-\infty}^{\infty} |m(t)|^q \rho_Q (m,t) \rm{d} \mit{m}$ ($q>0$), which for strong anomalous diffusion process has been conjectured to satisfy $\langle |m(t)|^q \rangle \sim t^{q \nu(q)}$ with $\nu(q)$ not a constant~\cite{Bilinear-1}. For the specific L\'evy walks model with $1<\gamma<2$ ($\gamma$ here is the power law exponent from the waiting time distribution $\phi (\tau) \sim \tau^{-1-\gamma}$ of the model, see~\cite{Bilinear,Bilinear-3} for details), it has been predicted that for the low order $q$, $\langle |m(t)|^q \rangle \sim t^{q /\gamma}$; while for high order $q$, $q \nu(q)=q+1-\gamma$~\cite{Bilinear,Bilinear-3}. As example, the result of $q \nu(q)$ versus $q$ in our case under $T=0.75$ is plotted in Fig.~\ref{Fig4}. As expected, the bi-linear scaling can be clearly verified. Here the fitting value for low order $q$ is $\nu(q) \simeq 0.63$, coincident with the prediction of $1/\gamma \simeq 0.60$ [$\gamma=1.66$ from Fig.~\ref{Fig3}(d)]. Such coincidence indicates that the dynamical scaling exponent $\gamma$ considered here might correspond to the power law exponent of the waiting time distribution in L\'evy walks model.

\begin{figure}
\vskip-.2cm \hskip-0.4cm
\includegraphics[width=8.cm]{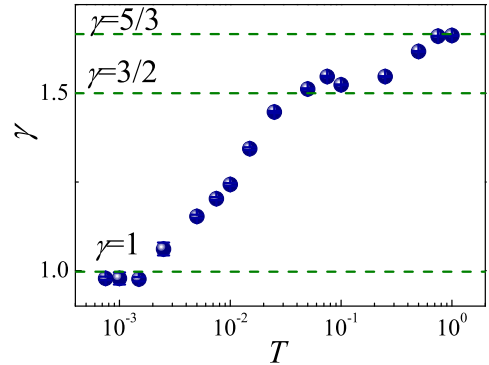}
\vskip-0.5cm \caption{\label{Fig5}(Color online) $\gamma$ versus $T$, the horizontal dashed lines, from bottom to top,
denote $\gamma=1$, $\gamma=3/2$ and $\gamma=5/3$, respectively.}
\end{figure}

Employing this scaling exponent $\gamma$ to connect anamolous heat transport is of great interest since from which one might get the time scaling exponent of the mean squared deviation of this heat diffusion process and thus connected to the system size dependent divergence exponent $\alpha$~\cite{Connect-1,Connect-2,Note_connection}. Due to this interest, figure~\ref{Fig5} further depicts the result of $\gamma$ versus $T$. Therein four data points are extracted
from Fig.~\ref{Fig3}, while others are obtained by performing the same scaling analysis. This result indicates that with the increase of $T$, $\gamma$ first remains constant at about $\gamma \simeq 1$, then follows temperature dependent behaviors in the intermediate range of $T$, finally seems to saturate at $\gamma=5/3$ for high temperatures. Interestingly, such temperature variation of $\gamma$ is similar to that shown in FPU-$\beta$ chains~\cite{Xiong-4}, where the nonlinearity dependence of $\gamma$ values crossover between different universality classes have been reported. As comparison to theories, we note that a recent theory~\cite{Hydrodynamics-2} suggested two universality classes of $\gamma$, i.e., $\gamma=3/2$ and $\gamma=5/3$ for the systems with symmetric and asymmetric interactions under zero and non-zero internal averaged pressure $\langle F \rangle$, respectively; however, for the special ST anharmonic system baring symmetric potential and $\langle F \rangle=0$ considered here, it seems that the prediction of $\gamma=3/2$ is not always valid. In fact, such non-universal scaling law has also been supported by some other theories~\cite{Daswell-1,Daswell-2,Daswell-3,Popkov} and numerical results~\cite{Hurtado}.
\section{Underlying mechanism}
Why can we see such anomalous temperature dependent heat transport? Is there any new properties after including the ST anharmonic interaction? To answer these questions, we here first study the momentum spread and then explore the properties of phonons' damping with temperatures.
\begin{figure}
\vskip-.2cm \hskip-0.4cm
\includegraphics[width=8.8cm]{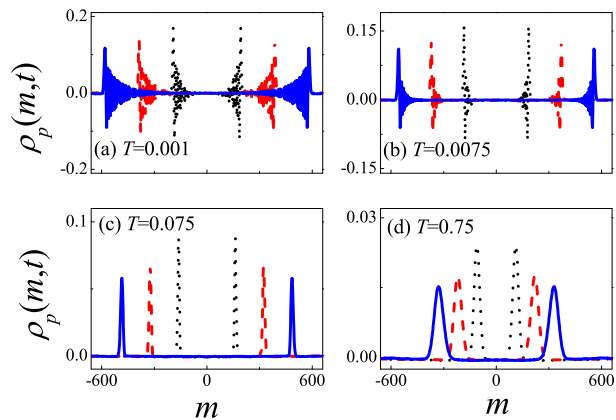}
\vskip-0.5cm  \caption{\label{Fig6}(Color online) Momentum spread $\rho_p(m,t)$ for three long time $t=200$ (dot), $t=400$ (dash) and $t=600$ (solid) under temperatures $T=0.001$ (a);
$T=0.0075$ (b); $T=0.075$ (c) and $T=0.75$ (d), respectively.}
\end{figure}
\subsection{Momentum spread}
The momentum spread contains useful information in understanding heat transport of momentum-coserving system. From the perspective of hydrodynamics theory, it may represent the diffusion of sound modes~\cite{Chen2013}. A recent work has attributed the observed normal heat transport in rotator systems to the diffusive behavior of momentum spread~\cite{YunyunLi}. This non-ballistic spread of momentum has also been found in a system with a double-well interparticle interaction under certain temperature ranges, where normal heat transport can be seen~\cite{Xiong-5}. For some integrable systems baring ballistic heat transport, based on a new concept of phonon random walks, the ballistic momentum spread is proved to be a quantum like wave function's real part~\cite{Ushape}. Therefore, ballistic (non-ballistic) momentum spread seems always the case for anomalous (normal) heat transport.

Figure~\ref{Fig6} depicts the results of momentum spread $\rho_p(m,t)$. As for comparison, three long time and four typical temperatures, the same as those in heat spread are considered. As can be seen, the momentum
spread also indicates interesting temperature dependent evidences: while at low temperatures there are some oscillations in the profiles of $\rho_p(m,t)$ [see Fig.~\ref{Fig6}(a)]; with the increase of $T$, such oscillations become less and less [see Fig.~\ref{Fig6}(b)], and eventually disappear [see Fig.~\ref{Fig6}(c)]; after that if one increases $T$ further, the front peaks begin to disperse [see Fig.~\ref{Fig6}(d)]. Thus this unusual change of $\rho_p(m,t)$ with temperatures but still following ballistic spreading may correspond to the anomalous temperature dependence of heat spread.

It was usually suggested that the velocity of the front peaks shown in the momentum spread just corresponds to the sound velocity $c$~\cite{NLi}. A recent theory~\cite{Hydrodynamics-2} proposed a general formula
\begin{equation} \label{SoundV}
c=\sqrt{ \frac{\frac{1}{2} T^{2} + \left\langle V + \left\langle
F\right\rangle \xi; V + \left\langle F\right\rangle \xi \right
\rangle } { \frac{1}{T} \left( \left\langle \xi; \xi \right \rangle
\left<V; V \right \rangle - \left\langle \xi; V \right\rangle ^{2}
\right)
 +\frac{1}{2} T \left\langle \xi; \xi \right\rangle } }
\end{equation}
which can predict the sound velocity for systems with any interparticle interaction. In formula (6), $V(\xi)$ is the interparticle potential, $\left\langle A; B \right\rangle$
denotes the covariance $\left\langle A B \right\rangle -
\left\langle A \right\rangle \left\langle B \right\rangle$ for any
two quantities $A$ and $B$, $\left\langle F \right\rangle$ is
the averaged pressure ($\left\langle F \right\rangle \equiv 0$ for symmetric potentials). It is thus worthwhile to check whether formula~\eqref{SoundV} is still valid here. For such purpose, we numerically measure the velocity of front peaks as shown in $\rho_p(m,t)$ for each temperature and compare the result with the prediction from formula~\eqref{SoundV}. To obtain the theoretical predictions, we insert the ST anharmonic potential [Eq.~\eqref{STPotential}] into formula~\eqref{SoundV} and calculate the ensemble average of each quantity
$\left\langle A \right\rangle$ by $\int_{-\infty}^{\infty} A
e^{-V(\xi)/T} \rm{d} \xi / \int_{-\infty}^{\infty}
\mit{e}^{-\mit{V}(\xi)/\mit{T}} \rm{d} \xi$.
\begin{figure}
\vskip-.2cm \hskip-0.4cm
\includegraphics[width=8.cm]{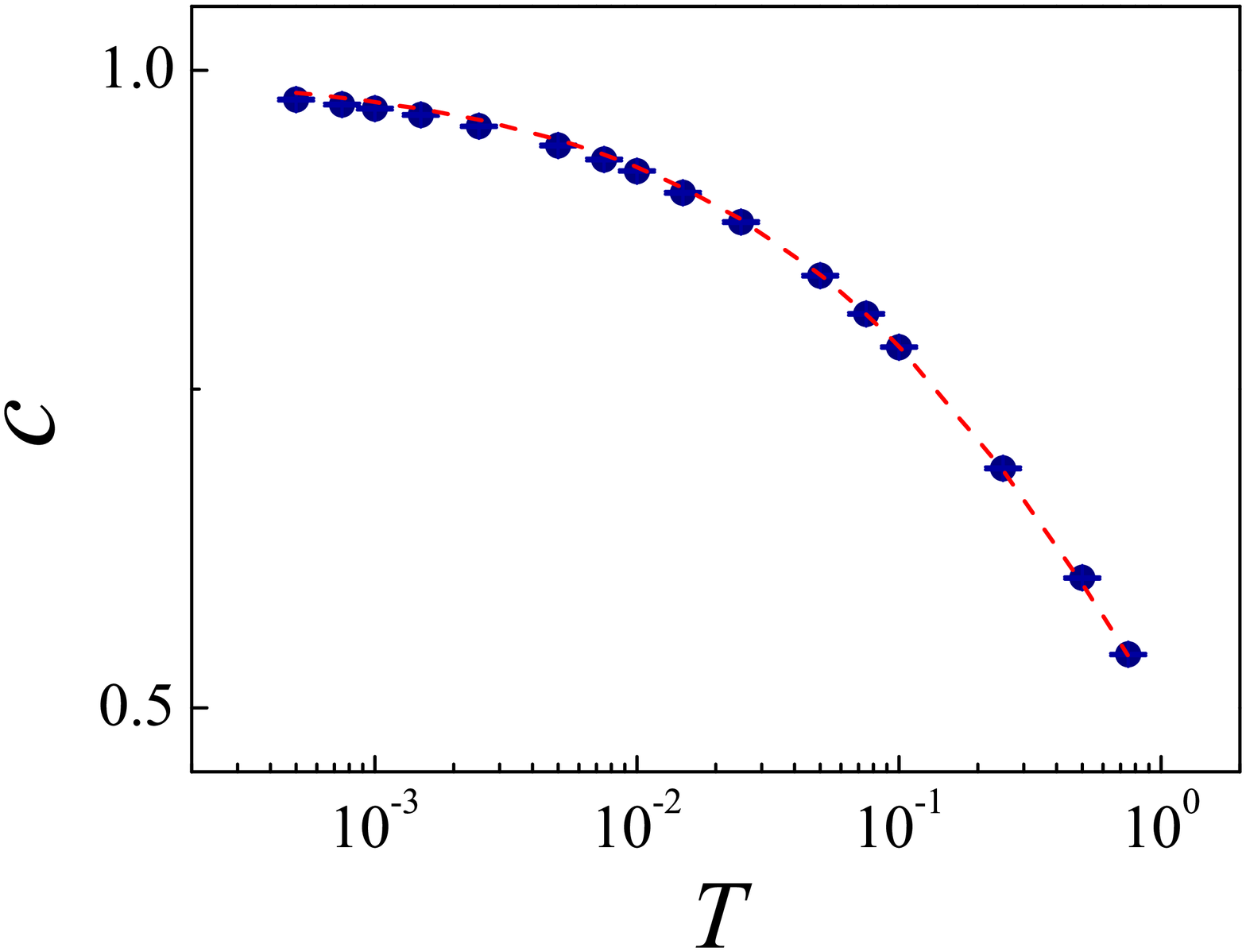}
\vskip-0.5cm \caption{\label{Fig7}(Color online) The sound velocity $c$ versus $T$,
where the dashed line denotes the predictions from formula~\eqref{SoundV}.}
\end{figure}

Figure~\ref{Fig7} shows the result of $c$ versus $T$. As can be seen, the numerical measurements match the predictions quite well, suggesting that indeed, the formula~\eqref{SoundV} can also be
validated to the systems with ST anharmonicity. More-importantly, both results indicate the decrease of
sound velocity with temperatures, which is clearly opposite to the results as shown in FPU-$\beta$ chain~\cite{NLi}. Thus, this may be a generic feature for systems with ST anharmonicity.
\subsection{Phonon spectrum}
Clearly, the decrease of sound velocity cannot be used to fully understand the temperature dependence of heat spread. We now turn to the analysis of system's phonon spectrum
$P(\omega)$, from which one may gain further insights. A quite recent work~\cite{Xiong-5} has suggested that, in addition to the non-ballistic behavior of momentum spread, a complete damping of phonons together with phonons' softening seem crucial to the observed normal heat transport ($\alpha=0$, satisfy Fourier's law). Therefore, it would be necessary to explore how phonons' damping and softening would play the roles here.

The phonons spectra $P(\omega)$ is
calculated by applying a frequency $\omega$ analysis of the particles' velocity $v(t)$ (see the appendix
of the review~\cite{BLiReview})
\begin{equation} \label{PW}
P(\omega)=\lim_{\tau \rightarrow \infty} \frac{1}{\tau}
\int_{0}^{\tau} v (t) \exp (- \rm{i} \mit{\omega} t) \rm{d}
\mit{t}.
\end{equation}
To be related to heat spread, this frequency analysis should be done at the corresponding equilibrium states under the same temperatures. For facilitating the computation, here we choose a chain of $L=200$ particles, then thermalize the chain to the focused temperature by Langevin
heat baths~\cite{Lepri_Report,Dhar_Report}, finally remove the heat baths and perform a frequency analysis of $v(t)$ following Eq.~\eqref{PW}. This should also be done by starting from properly assigned initial random states for several times.

\begin{figure}
\vskip-.2cm \hskip-0.2cm
\includegraphics[width=8.8cm]{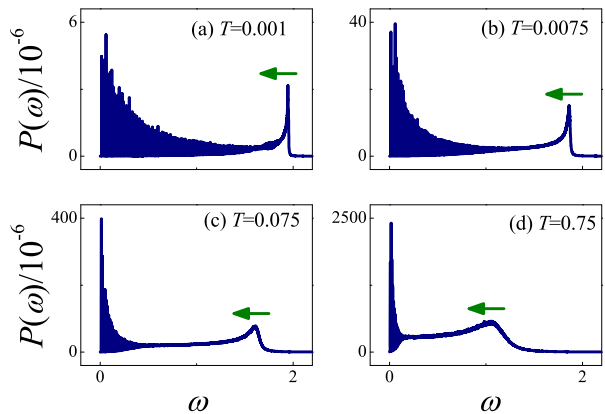}
\vskip-0.5cm \caption{\label{Fig8}(Color online) Phonons' spectrum
$P(\omega)$ for different temperatures: (a) $T=0.001$; (b) $T=0.0075$; (c) $T=0.075$; (d)
$T=0.75$, respectively.}
\end{figure}
\begin{figure}
\vskip-.2cm \hskip-0.2cm
\includegraphics[width=8.8cm]{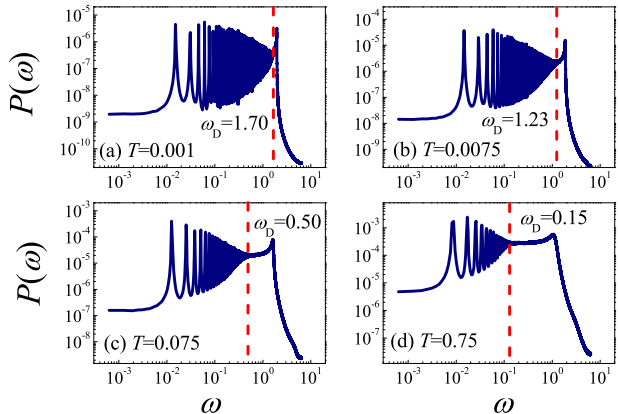}
\vskip-0.5cm  \caption{\label{Fig9}(Color online) A log-log plot of Fig.~\ref{Fig8}, where the dotted lines denote the frequencies $\omega_{\rm{D}}$ below which phonons are damped very weakly.}
\end{figure}
\begin{figure}
\vskip-.2cm \hskip-0.2cm
\includegraphics[width=8.cm]{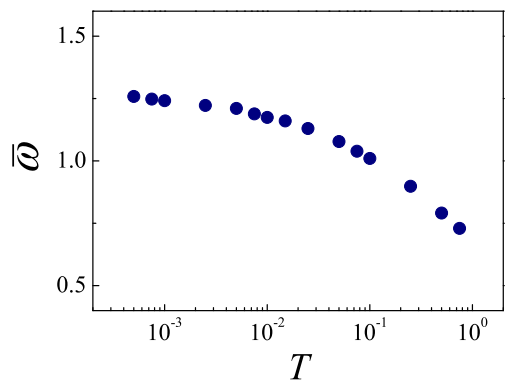}
\vskip-0.5cm  \caption{\label{Fig10}(Color online) The averaged frequency
$\bar{\omega}$ of phonons shown in $P(\omega)$ versus temperature.}
\end{figure}

Figures~\ref{Fig8} and~\ref{Fig9} depict the results of phonons spectrum $P(\omega)$ for different temperatures. Two key points can be revealed from the results. First, with the increase of $T$, opposite to the systems with HT anharmonicity (hardening phonons), phonons here tend to become ``softer" since $P(\omega)$ walks towards the direction of low frequency. This can be captured from the locations of the peaks in the high frequency parts (see Fig.~\ref{Fig10}). To more clearly characterize this phonons' softening process, one can measure the averaged frequency $\bar{\omega}$ of phonons by defining $\bar{\omega}={\int_{0}^{\infty} P (\omega) \omega \rm{d}
\mit{\omega}} / {\int_{0}^{\infty} P (\omega) \rm{d} \mit{\omega}}$. As complementary we plot $\bar{\omega}$ versus $T$ in Fig.~\ref{Fig10}, from which a monotonous decrease of $\bar{\omega}$ different from the non-monotonous case as shown in Ref.~\cite{Xiong-5} can be clearly seen. This seems to suggest that only a monotonous phonons' softening process is inadequate to induce normal heat transport. To realize the normal transport behavior, very high temperatures or other factors might be necessary to take into account, which needs further efforts of investigations.

Let us finally turn to the results of phonons' damping. Since if they are still called phonons, in their power spectrum at associated frequencies, there should be some oscillations. Indeed, Ref.~\cite{Xiong-5} has suggested that a complete absence of oscillations implies the normal heat transport. With this in mind one then can employ Fig.~\ref{Fig9} to explore this phonons' damping process. As can be seen, with the increase of $T$, the damping first originates from the high frequency parts [see Figs.~\ref{Fig9}(a) and (b)] and then quickly towards the low ones [see Fig.~\ref{Fig9}(c)]. However,
such quick damping process cannot last forever if one further increases the temperature [see Fig.~\ref{Fig9}(d)], eventually, a power spectrum of phonons very similar to that shown in FPU-$\beta$ systems with HT anharmonicity (under high nonlinearity)~\cite{Xiong-4} can be seen. Note that here we use the critical frequency of $\omega_{\rm{D}}$ (below which phonons are damped very weakly) to characterize this phonons' damping process. It is worthwhile to recognize that this incomplete damping process may correspond to the L\'{e}vy walks densities observed in heat spread under high temperatures. It also suggests that both ST and HT anharmonicity can only lead to an incomplete damping process of phonons, thus generally, an universal anomalous heat transport with certain scaling exponents could be observed in nonlinear systems with only ST or HT anharmonicity.
\section{Summary}
In summary, we have studied the temperature dependent heat transport behavior in a 1D system when the ST interparticle interaction is considered. We have found that by increasing the temperature, including the ST anharmonicity can induce some opposite effects to its counterpart systems where the interparticle interactions are HT, such as that \emph{monotonously} softening the phonons and decreasing the sound velocity. However, such unusual properties are still inadequate to lead to normal heat transport. A analysis of phonons spectra indicates an incomplete damping process of phonons, especially those at low frequencies. This property of spectra is similar to that shown in FPU-$\beta$ systems with HT anharmonicity. Our results thus suggest that both ST and HT anharmonicity will eventually lead to a general super-diffusive heat transport behavior, therefore further supporting the conjecture that even strong nonlinearity (here only the deterministic dynamics are considered, which may lead to chaos), either from HT anharmoncity or from ST anharmoncity is, neither a sufficient nor a necessary condition for the validity of Fourier's law~\cite{Casati,PRL1997}.
\begin{acknowledgments}
This work was supported by the NNSF (Grant No. 11575046) of China; the training plan for Distinguished Young researchers of Fujian provincial department of education; the Qishan scholar research fund of Fuzhou university.
\end{acknowledgments}


\begin{thebibliography}{99}
\bibitem{Note_connection} In L\'evy walks theory there is a generic relation $\alpha=2-\gamma$, which is obtained by simultaneously adopting $\alpha=\mu-1$ and $\mu=3-\gamma$, where $\mu$ is the time scaling exponent of the displacement's second momentum from the energy fluctuations spreading: $\sigma^2(t) \sim t^{\mu}$, see~\cite{Connect-1} for details.
\end{thebibliography}
\end{document}